\documentclass[journal]{new-aiaa}
\usepackage[utf8]{inputenc}

\usepackage{graphicx}
\usepackage{amsmath,bm}
\usepackage[version=4]{mhchem}
\usepackage{siunitx}
\usepackage{longtable,tabularx}
\usepackage{hyperref}
\usepackage{subcaption}
\usepackage{tabularx}
\usepackage{booktabs}
\renewcommand{\arraystretch}{1.25}

\title{Multifidelity Graph Neural Networks for 3D Flow Prediction in Aircraft Design Optimization}

\author{
Mark Sperry\footnote{PhD Candidate, Department of Mechanical and Aerospace Engineering. AIAA Student Member}\\
University of California San Diego, 9500 Gilman Dr, La Jolla, CA 92093
\and
Dean E. Bryson\footnote{Senior Research Aerospace Engineer, Air Warfare Directorate. AIAA Associate Fellow}\\
Air Force Research Laboratory, Wright-Patterson Air Force Base, OH 45433
\and
John T. Hwang\footnote{Associate Professor, Department of Mechanical and Aerospace Engineering. AIAA Senior Member}\\
University of California San Diego, 9500 Gilman Dr, La Jolla, CA 92093
}

\begin{document}

\maketitle
 
\begin{abstract}
A major bottleneck in high-fidelity, adjoint-based aircraft multidisciplinary design optimization (MDO) is the cost of repeatedly evaluating computational fluid dynamics (CFD) models.
Modern scientific machine learning models for aerodynamics are computationally efficient, but often focus on 2D prediction, are single-fidelity, or are not considered for use in gradient-based optimization. 
We present a multifidelity graph neural network (GNN) method for 3D flow prediction and apply it to RANS analysis around a blended-wing-body aircraft concept.
The method embeds a low-fidelity solver into the model workflow to provide a base prediction that the GNN corrects towards RANS flow fields.
Trained on 110 RANS simulations across a 131-dimensional parameter space, the panel-guided and Euler-guided GNNs achieve aerodynamic coefficient relative $L_2$ test errors below 8.6\% and 5.6\%, respectively, while evaluating 100 and 9 times faster than RANS CFD.
Compared with the single-fidelity GNN baseline, the panel-guided and Euler-guided GNNs reduce volume-field prediction errors by approximately 14--25\% and 29--48\%, respectively.
Finally, we apply the single-fidelity and panel-guided GNN models to gradient-based MDO with 152 design variables and 13 constraints.
Post-optimality RANS re-evaluations yield lift and drag errors of 6.4\% and 17.5\% for the panel-guided GNN, compared with 30.3\% and 45.7\% for the single-fidelity GNN.
\end{abstract}

\section{Introduction} \label{section:intro}
Multidisciplinary design optimization (MDO) is an engineering design methodology that applies numerical optimization to improve system performance or efficiency using models that consider multiple disciplines.
Solving large-scale MDO problems can take several hours of wall-clock time \cite{mdo_long}, or even days when high-fidelity models such as computational fluid dynamics (CFD) models are involved.

To mitigate this issue, engineers commonly rely on low-fidelity models with simplified physics or data-driven surrogate models.
For aerodynamic modeling in aircraft design, low-fidelity models often come in the form of vortex lattice methods or panel methods, which are computationally efficient and predict surface field quantities.
These approaches neglect viscous, turbulent, and compressible flow effects.
Traditional surrogate modeling techniques used in aerodynamic optimization, such as Kriging and radial basis function methods, have commonly focused on predicting integrated quantities such as lift and drag directly from CFD data.
However, full-field predictions are often necessary to support multidisciplinary coupling, such as for aerostructural analysis.

Recent advances in machine learning have led to growing interest in surrogate models for aerodynamic field prediction \cite{bonnet2023airfranshighfidelitycomputational}.
Graph neural networks (GNNs) are a natural choice for CFD surrogate modeling because they can operate directly on unstructured discretizations of the geometry.
Early studies demonstrated the use of graph-based message passing to learn the dynamics of physical systems \cite{2018gnns,sanchez2020learning}.
MeshGraphNets extended this work to unstructured meshes \cite{pfaff2020learning}, while X-MeshGraphNet addressed issues associated with scalability and long-range interactions \cite{xmeshgraphnet}.
For aerodynamics, Hines and Bekemeyer used GNNs to predict large-scale aircraft surface pressure distributions, demonstrating the applicability of graph-based surrogate models for industrial aerodynamic applications \cite{HINES2023108268}.
Roznowicz et al. demonstrated large-scale graph-based surrogate modeling for full 3D volumetric aerodynamic flow fields \cite{roznowicz2024large}.
Other works have explored diffusion-based graph models for learning distributions of fluid states \cite{lino2025learning} and physics-informed GNNs that incorporate governing equations and boundary conditions \cite{PENG2024117551}.

In addition to GNNs, other architectures have also been explored for flow-field prediction.
Fourier neural operators learn in Fourier space to predict PDE solutions, showing success in fluids applications \cite{li2020fourier}.
Geometry-informed neural operators extend this by allowing Fourier-based operators to be evaluated on irregular geometry using graph neural operators \cite{gino}.
Deep operator networks have also been explored for geometry-dependent compressible aerodynamic flow prediction, including those in hypersonic and supersonic regimes \cite{fdeeponet}.
Implicit neural representations have been used to predict aerodynamic fields and were shown to outperform MeshGraphNets on some aerodynamic prediction tasks \cite{catalani2024neural}.
Recent studies have also investigated full 3D flow-field prediction using structured-grid architectures such as Flow3DNet \cite{zuo2025flow3dnet} and multiscale point-cloud-based approaches such as DoMINO \cite{domino}.
Other approaches include attention-based models for high-speed flow prediction \cite{highspeed} and transformer-based neural PDE operators for general geometries and large-scale applications, such as Transolver \cite{transolver} and AB-UPT \cite{abupt}.

These studies show that recent machine learning models are able to predict volumetric aerodynamic states around complex geometries.
Most of these approaches use high-fidelity CFD data for training and rely on the trained network alone at inference time.
An alternative approach is hybrid modeling, which aims to combine inexpensive, low-fidelity models with high-fidelity data.
In multifidelity modeling, models leverage information from different levels of fidelity to improve predictive accuracy.
Discrepancy modeling involves learning a surrogate to correct a lower-fidelity model to the desired higher-fidelity behavior \cite{Kennedy2000PredictingTO}.

Prior work has shown that multifidelity surrogate models can improve prediction accuracy compared with models trained only on high-fidelity data, especially when the amount of high-fidelity data is limited.
Hybrid approaches have combined differentiable solvers with neural networks, such as the architecture proposed by Belbute-Peres et al. \cite{belbute2020combining}.
Other approaches learn discrepancies to correct low-fidelity predictions, as demonstrated by VortexNet for surface pressure prediction using a vortex lattice method and GNNs \cite{vortex1}.
Multifidelity GNN approaches have also been developed for PDE modeling and turbomachinery flow-field prediction \cite{multisurrogate,multi2}.
These ideas have also been investigated for aerodynamic design optimization, including GNN-based optimization \cite{vortex2} and multifidelity deep learning approaches that combine low-fidelity and high-fidelity CFD data for shape optimization \cite{ZHANG2021113485, aobo}.

Few studies have investigated the use of low-fidelity solver predictions as inputs or as a correction baseline for full 3D aerodynamic flow-field prediction.
Additionally, the integration of solver-guided full-field surrogates into differentiable workflows for gradient-based aircraft design optimization remains limited.

In this work, we develop and investigate a multifidelity GNN method for 3D aerodynamic flow prediction.
This method uses a low-fidelity physics-based model as a base and applies a correction from a GNN trained on full-field data generated by the high-fidelity model that the multifidelity model aims to emulate.
We apply the method to Reynolds-averaged Navier--Stokes (RANS) analysis of a blended-wing-body (BWB) aircraft concept with a panel method and an Euler CFD solver as the low-fidelity models.
We compare the panel-guided and Euler-guided GNNs with the single-fidelity GNN baseline in terms of prediction accuracy and evaluation time.
Finally, we apply the single-fidelity and panel-guided GNNs to two gradient-based optimization problems, one for drag minimization over camber and twist parameters and one for large-scale MDO with 152 design variables and 13 constraints.

\section{Multifidelity Graph Neural Networks for 3D Aerodynamic Flow Prediction} 
\label{section:methodology}

This section presents the multifidelity GNN methodology for 3D aerodynamic flow prediction.
We first introduce the multifidelity formulation and overall model workflow.
We then describe the multiscale GNN architecture, including graph and input feature construction, then the training approach.

\subsection{Method Overview and Formulation}
\label{section:methodology:overview}

\begin{figure}[htbp]
    \centering
    \includegraphics[width=0.95\textwidth]{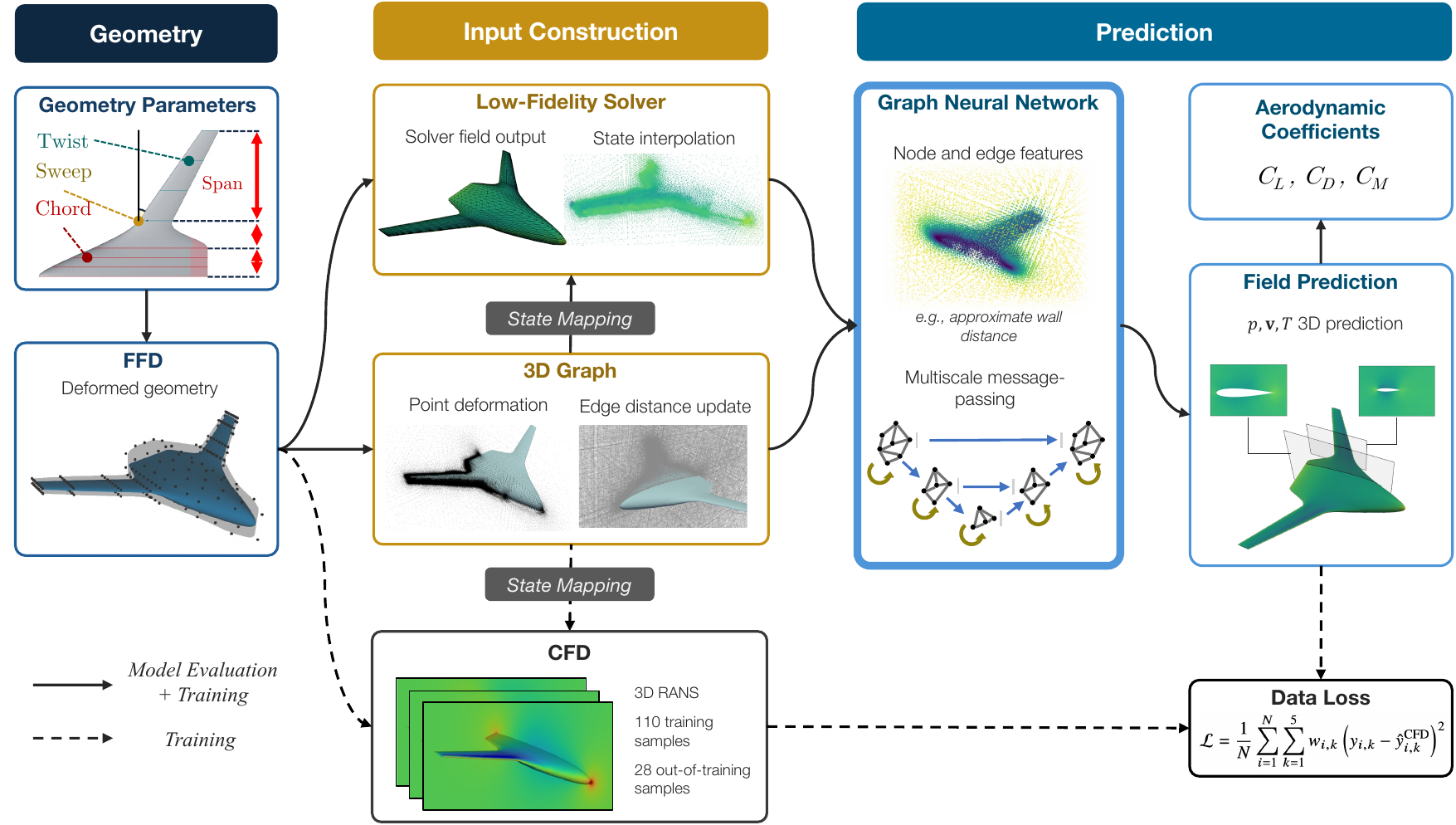}
    \caption{An overview of the training and evaluation workflow for the multifidelity GNN. The single-fidelity GNN shares the same architecture as the multifidelity GNN, but without the low-fidelity input step. Solid arrows show the evaluation steps used during online optimization, while dashed arrows show the additional steps required for training.}
    \label{fig:workflow}
\end{figure}

The overall workflow is shown in Fig.~\ref{fig:workflow}.
First, geometry parameters define the aircraft surface.
The volume graph on which the GNN is evaluated is constructed from a volumetric CFD mesh, and changes to the aircraft surface are propagated to the graph through differentiable mesh deformation.
For the multifidelity models, a low-fidelity solution field is mapped onto the graph and is used as the base prediction.
The GNN then predicts a correction to this field.
Aerodynamic quantities such as lift, drag, and pitching moment can then be computed from the predicted field.
During training, 3D RANS CFD field solutions are used as the training data.

The multifidelity GNN uses an approximate low-fidelity field prediction as the flow inputs at each node and predicts a correction to this state. The resulting normalized flow prediction is given by
\begin{equation} \label{eq:mf}
y_i^{\mathrm{MF}} = \hat{y}_i^{\mathrm{LF}} + \bar{y}_i(\hat{\mathbf{y}}^{\mathrm{LF}}),
\end{equation}
where $\hat{y}_i^{\mathrm{LF}} \in \mathbb{R}^5$ denotes the normalized low-fidelity state mapped to node $i$, and $\bar{y}_i$ is the GNN output. The GNN output depends on 
the low-fidelity field $\hat{\mathbf{y}}^{\mathrm{LF}}$, 
since the low-fidelity states are included in the input 
node features, as described in Section \ref{section:methodology:gnn}. In contrast, the standard single-fidelity GNN uses normalized freestream values as the flow inputs at each node and predicts the normalized RANS flow state directly:
\begin{equation} \label{eq:sf}
y_i^{\mathrm{SF}} = \bar{y}_i.
\end{equation}

Aside from the source of the flow inputs and the differences between Eqs.~(\ref{eq:mf}) and~(\ref{eq:sf}), the multifidelity and single-fidelity models use the same GNN architecture.

To enable gradient-based design optimization, derivatives of the predicted flow field and aerodynamic quantities with respect to the design variables can be obtained by propagating derivatives through the workflow.
For the multifidelity models, this requires additional derivatives through the low-fidelity solver.

\subsection{Multiscale Graph Neural Network Architecture}
\label{section:methodology:gnn}

\begin{figure}[htbp]
    \centering
    \includegraphics[width=0.95\textwidth]{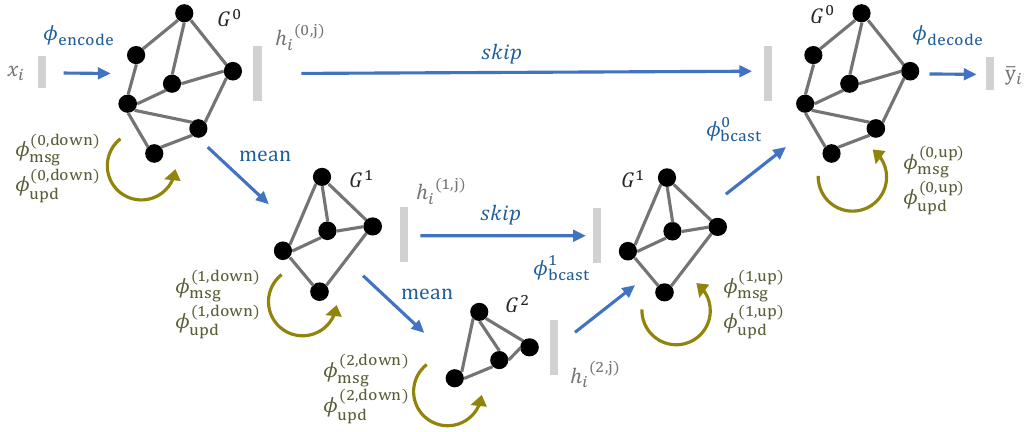}
    \caption{The multiscale GNN architecture with three graph levels. The fine graph is constructed directly from the CFD mesh while the coarser graphs are constructed by clustering nearby nodes.}
    \label{fig:unet}
\end{figure}

The GNN used in the multifidelity model follows a custom U-shaped GNN multiscale architecture (Fig.~\ref{fig:unet}) based on the Graph U-Net idea \cite{graph_unet}.
We use the dual graph of the CFD volume mesh to represent the fluid domain around the BWB.
Specifically, volume mesh cell centers are treated as graph nodes and shared faces define edges between the nodes.
Since volume cell centers do not exist on the wall, farfield, and symmetry faces, we build additional graph nodes on their respective face centers to encode boundary condition information into the graph.
Edges are connected between these boundary nodes and their corresponding adjacent cell volume nodes.
We denote this graph by
\begin{equation} \label{eq:graph}
    G^{0}=(V^0,E^0),
\end{equation}
where each node in $V^0$ corresponds to a mesh cell center or a boundary face center, and edges connect nodes with shared faces in the mesh. 

To create a multiscale hierarchy with $L$ graph levels, the coarser graphs $G^\ell = (V^\ell,E^\ell)$, $\ell=1,\ldots,L-1$, are constructed by successively applying graph clustering passes starting from $G^0$.
In each pass, candidate edges are first sorted by length, and nodes connected by these edges are iteratively merged if the edge lengths are below a fixed threshold.
To keep boundary information, edges connected to boundary nodes are excluded from clustering.
Each coarsening pass builds a fine-node to coarse-node map which is stored for downsampling and upsampling between graph levels.
Coordinates for the coarsened nodes are taken as the average of the coordinates of the fine nodes assigned to that cluster.

\begin{table}[t]
\centering
\caption{Graph neural network input and output features.}
\label{tab:features}
\begin{tabular}{lll}
\toprule
Category & Feature & Type \\
\midrule
Node input
    & Normalized freestream/approximate states
    & $\mathbb{R}^{5}$ \\
Node input
    & Node-type indicator
    & $\{0,1\}^{4}$ \\
Node input
    & Approximate wall distance
    & $\mathbb{R}$ \\
Node input
    & Approximate nearest wall direction
    & $\mathbb{R}^{3}$ \\
Extra wall-node input
    & Approximate wall face area
    & $\mathbb{R}$ \\
Extra wall-node input
    & Surface unit normal
    & $\mathbb{R}^{3}$ \\
\midrule
Edge input
    & Edge direction
    & $\mathbb{R}^{3}$ \\
Edge input
    & Edge length
    & $\mathbb{R}$ \\
Edge input
    & Edge-type indicator
    & $\{0,1\}^{4}$ \\
\midrule
Volume-node output
    & Normalized primitive flow variables
    & $\mathbb{R}^{5}$ \\
\bottomrule
\end{tabular}
\end{table}

Inputs to the network are passed in as features for each node $i\in V^0$ and edge $(i,j)\in E^\ell$ as summarized in Table~\ref{tab:features}.
The base input node feature vector $x_i\in \mathbb{R}^{13}$ includes five normalized flow state values ($p$, $\mathbf{v}$, and $T$), taken from the freestream conditions for the single-fidelity model or the mapped low-fidelity field for the multifidelity models, as described in the preceding section.
The remaining features consist of an approximate wall-distance value, an approximate unit vector pointing to the nearest wall node, and a one-hot indicator for whether the node is part of the fluid domain, wall, farfield, or symmetry plane.
Nodes located on the wall additionally have the approximate area of the corresponding wall face and its unit normal, resulting in the feature vector $x_i^{\mathrm{wall}} \in \mathbb{R}^{17}$.
Edge features $e_{ij} \in \mathbb{R}^{8}$ include the edge's unit direction, length and corresponding one-hot type indicators and are built for all edge levels.
Because message-passing is performed in both directions, edge features are constructed for both orientations, where the unit direction vector is reversed.

The node features on $V^0$ are encoded into $d_h$-dimensional latent vectors using separate encoder networks for wall and non-wall nodes.
The initial latent vector is then
\begin{equation}
h_i^{(0,0)}
=
\begin{cases}
\phi_{\mathrm{encode}}^{\mathrm{wall}}
    \left(x_i^{\mathrm{wall}}\right),
    & i\in V_{\mathrm{wall}}^0, \\[3pt]
\phi_{\mathrm{encode}}^{\mathrm{base}}
    \left(x_i\right),
    & i\notin V_{\mathrm{wall}}^0.
\end{cases}
\end{equation}
Here,
$\phi_{\mathrm{encode}}^{\mathrm{base}}:
\mathbb{R}^{13}\rightarrow\mathbb{R}^{d_h}$
and
$\phi_{\mathrm{encode}}^{\mathrm{wall}}:
\mathbb{R}^{17}\rightarrow\mathbb{R}^{d_h}$
are independently parameterized MLPs.

The network alternates between message-passing iterations and pooling during downsampling, followed by learned unpooling and additional message-passing steps during upsampling.
We denote the downsampling/upsampling direction by $s\in\{\mathrm{down},\mathrm{up}\}$.
The downsampling path traverses graph levels from $0$ to $L-1$, while the upsampling path traverses levels from $L-2$ back to $0$.
After message-passing at a fine level on the downsampling path, the latent vectors are averaged within each cluster to initialize the next coarser graph.

At graph level $\ell$ and sampling direction $s$, a directional message along an edge $(i,j)\in E^\ell$ at message-passing iteration $t$ is computed as
\begin{equation}
    m_{ij}^{(\ell,t)}
    =\phi_{\mathrm{msg}}^{(\ell,s)}
    \Big(
    \big[
    h_i^{(\ell,t)},\;
    h_j^{(\ell,t)},\;
    e_{ij}^{\ell}
    \big]
    \Big),
\end{equation}
where each $\phi_{\mathrm{msg}}^{(\ell,s)}: \mathbb{R}^{2d_h+8} \to \mathbb{R}^{d_h}$ is a learned MLP specific to that stage.
Because message-passing is performed in both edge directions, the aggregated message at a node $i$ is
\begin{equation}
    \bar m_i^{(\ell,t)}
=
\frac{1}{\deg(i)}
\sum_{j\in\mathcal N(i)}
m_{ji}^{(\ell,t)},
\end{equation}
where $\mathcal N(i)$ and $\deg(i)$ are the neighboring nodes and the number of neighbors for node $i$.
The global latent vector is the mean over the interior volume nodes at that level,
\begin{equation}
g^{(\ell,t)}
=
\frac{1}{|V_{\mathrm{int}}^\ell|}
\sum_{k\in V_{\mathrm{int}}^\ell}
h_k^{(\ell,t)}.
\end{equation}
The node latent vectors are then updated through a residual network:
\begin{equation}
h_i^{(\ell,t+1)}
=
h_i^{(\ell,t)}
+
\phi_{\mathrm{upd}}^{(\ell,s)}
\Big(
\big[
h_i^{(\ell,t)},\;
\bar m_i^{(\ell,t)},\;
g^{(\ell,t)}
\big]
\Big),
\end{equation}
where each $\phi_{\mathrm{upd}}^{(\ell,s)}: \mathbb{R}^{3d_h} \to \mathbb{R}^{d_h}$ is also specific to stage $(\ell,s)$.
Within a stage, the same message and update parameters are reused for every message-passing iteration and for both edge orientations.
Parameters are not shared between graph levels; in particular, the downsampling and upsampling stages at the same graph level use different MLPs.

During the unpooling stage, coarse node latent vectors are upsampled back to fine latent vectors through a learned network based on the coarse node latent vector and saved skip connections:  
\begin{equation}
\tilde h_i^{(\ell)}
=
h_i^{(\ell,\mathrm{skip})}
+
\phi_{\mathrm{bcast}}^{\ell}
\Big(
\big[
h_i^{(\ell,\mathrm{skip})},\;
h_{C(i, \ell+1)}^{(\ell+1)}
\big]
\Big),
\end{equation}
where each $\phi_{\mathrm{bcast}}^{\ell}: \mathbb{R}^{2d_h} \to \mathbb{R}^{d_h}$ is an independently parameterized MLP for graph level $\ell=0,\ldots,L-2$, and $C(i, \ell+1)$ is the corresponding clustered node in $V^{\ell+1}$ associated with node $i\in V^{\ell}$.
Finally, the outputs for interior volume nodes are decoded using a final decoder network on $G^{0}$:
\begin{equation}
    \bar{y}_i=\phi_{\mathrm{decode}}(h_i^{(0,\mathrm{out})})\in\mathbb{R}^5,
    \qquad i\in V_{\mathrm{int}}^0,
\end{equation}
where $\bar{y}_i$ corresponds to the five outputs associated with $p$, $\mathbf{v}$, and $T$.

\subsection{Supervised Field Training}
\label{section:methodology:training}

The network is trained in a supervised manner to predict the five primitive flow variables on the graph for each training geometry.
All target variables are normalized using the mean and standard deviation from the training set.
For a single training geometry, the loss is defined as the sum of the mean-squared error between the predicted and normalized states over all interior graph nodes across the five states:

\begin{equation}
\mathcal{L}
=
\frac{1}{N}
\sum_{i=1}^{N}
\sum_{k=1}^{5}
w_{i,k}
\left(
y_{i,k} - \hat{y}_{i,k}^{\mathrm{CFD}}
\right)^2
\label{eq:loss}
\end{equation}
where \(N\) is the total number of graph nodes in the fluid domain, \(k\) indexes the five
predicted flow variables, \({y}_{i,k}\) is the predicted normalized
value of state \(k\) at node \(i\), and \(\hat{y}_{i,k}^{\mathrm{CFD}}\) is the corresponding normalized CFD target value.
The weights $w_{i,k}$ allow greater emphasis to be placed on specific flow variables or regions of the fluid domain during training.
Although scalar forces are available, they are not included in the loss function and the model is only trained on state values so as not to overemphasize integrated quantities.

\section{Numerical Setup and Model Implementation}  
\label{section:setup}

This section describes the numerical setup and implementation of the GNN models for the blended-wing-body problem (Section \ref{section:setup:bwb_rans_reference}), with RANS CFD as the high-fidelity reference model.
The single-fidelity baseline is called RANS-GNN, and the two multifidelity variants are called panel-guided RANS-GNN (PG-RANS-GNN) and Euler-guided RANS-GNN (EG-RANS-GNN), which use a panel method and an Euler CFD solver, respectively.
These models are used for the performance analysis in Section \ref{section:results}, while the RANS-GNN and PG-RANS-GNN are used for design optimization in Section \ref{section:optimization}.
The GNN model implementations, training workflows, and optimization scripts are available in the open-source \texttt{AeroDML} repository \footnote{\url{https://github.com/LSDOlab/AeroDML}}.

\subsection{BWB Problem and RANS Reference Dataset}
\label{section:setup:bwb_rans_reference}

\begin{table}[t]
\centering
\small
\caption{Independent BWB geometry and flow condition variables and bounds used for Latin hypercube sampling.}
\label{tab:geometry_parameters}
\begin{tabular}{lcc}
\hline
\textbf{Variable} & \textbf{No. of parameters} & \textbf{Sampling range} \\
\hline
centerbody chord ratios
    & 4
    & $0.9$--$1.1$ \\
wing-root longitudinal position
    & 1
    & $13.2$--$22.2\ \mathrm{m}$ \\
wing sweep angle
    & 1
    & $15^\circ$--$42^\circ$ \\
wing half-span
    & 1
    & $8$--$20\ \mathrm{m}$ \\
centerbody half-span
    & 1
    & $3$--$7\ \mathrm{m}$ \\
transition half-span
    & 1
    & $3.5$--$7\ \mathrm{m}$ \\
wing chord-stretch coefficients
    & 4
    & $\pm(3.0,\,2.5,\,2.0,\,1.5)\ \mathrm{m}$ \\
wing twist coefficients
    & 4
    & $\pm 5^\circ$ \\
airfoil thickness (B-spline parameters)
    & $8 \times 8$
    & $\pm 5\%$ of local baseline thickness \\
airfoil camber (B-spline parameters)
    & $8 \times 6$
    & $\pm 0.03c$ \\
elevator deflection
    & 1
    & $0^\circ$--$7^\circ$ \\
pitch
    & 1
    & $\pm5^\circ$ \\
\hline
\textbf{Total}
    & \textbf{131}
    & \\
\hline
\end{tabular}
\end{table}

\begin{figure}[htbp]
    \centering
    \begin{subfigure}[b]{0.50\textwidth}
        \centering
        \includegraphics[width=\linewidth,trim=25bp 55bp 0bp 50bp,clip]{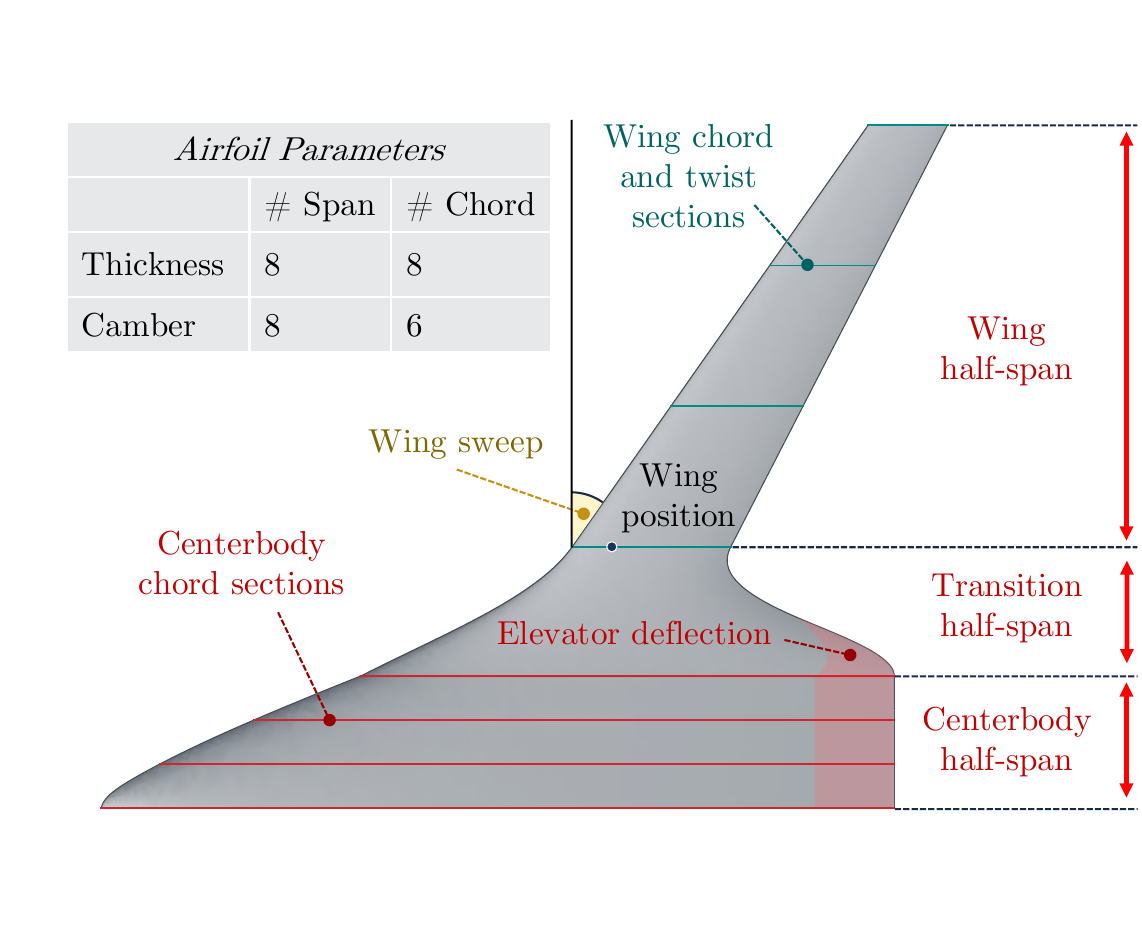}
        \caption{130 geometry parameters and their locations on the baseline geometry.}
        \label{fig:parameterization}
    \end{subfigure}\hfill
    \begin{subfigure}[b]{0.47\textwidth}
        \centering
        \includegraphics[width=\linewidth,trim=240bp 15bp 140bp 0bp,clip]{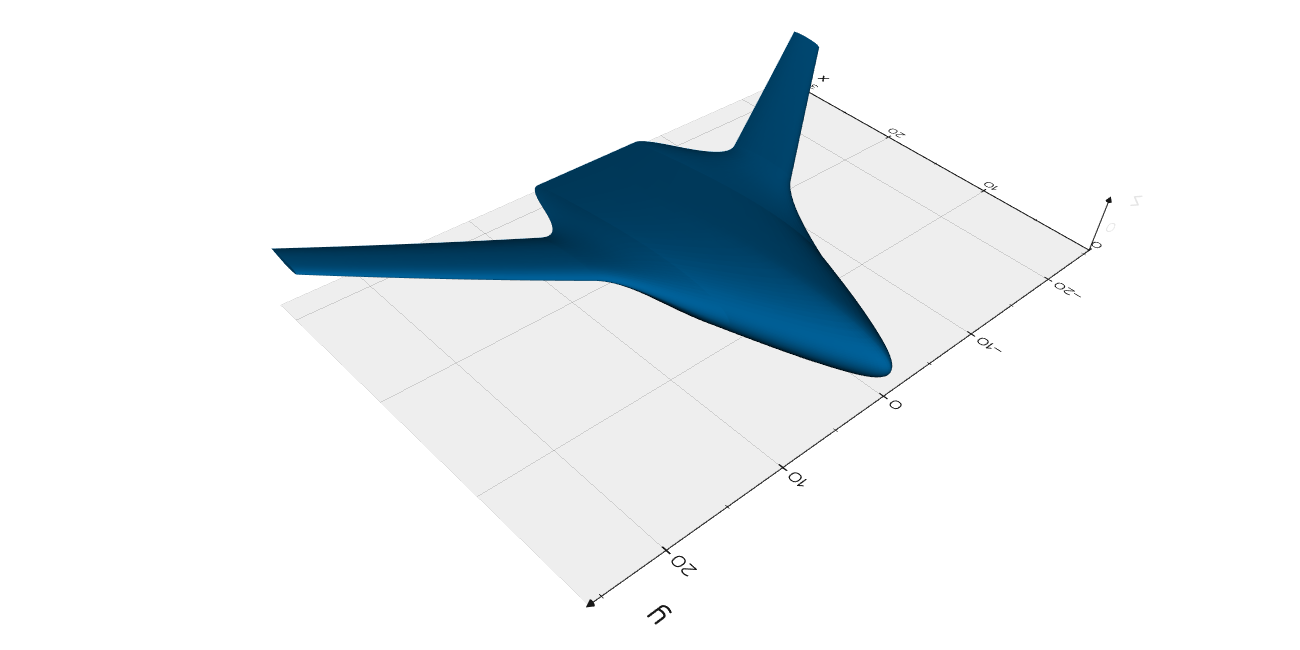}
        \caption{Baseline BWB geometry.}
        \label{fig:geometry}
    \end{subfigure}
    \caption{BWB geometry parameterization and baseline configuration.}
    \label{fig:bwb_configuration}
\end{figure}

We consider the problem of predicting the 3D flow field around a blended-wing-body aircraft concept at a nominal cruise condition.
The geometry, originally created in OpenVSP, is parameterized by a set of planform, twist and airfoil shape variables illustrated in Fig.~\ref{fig:parameterization} and summarized in Table~\ref{tab:geometry_parameters} along with the sampling ranges used for data generation.
The airfoil thickness and camber variables control displacements of the B-spline control points in the free-form deformation block.
The baseline configuration, as shown in Fig.~\ref{fig:geometry}, uses a symmetric airfoil in the centerbody, which blends into a transonic airfoil on the outer wing.
The parameterization is implemented using \texttt{lsdo\_geo}\footnote{\url{https://github.com/LSDOlab/lsdo_geo}} \cite{AndrewGeoParam}, a differentiable geometry software package developed for multidisciplinary design optimization (MDO).
All cases are evaluated at Mach 0.75 and an altitude of 30,000 ft.

The high-fidelity dataset is generated using steady RANS CFD simulations with the open-source \texttt{OpenFOAM} software.
Each case is evaluated on a fixed-topology 3D volume mesh with approximately 639k cells representing half of the fluid domain.
Boundary conditions include a no-slip wall condition on the aircraft surface, farfield conditions on the outer domain faces, and a symmetry condition along the aircraft centerbody plane.
Turbulence is modeled using the Spalart--Allmaras model.
A total of 138 converged CFD cases are generated using Latin hypercube sampling of the parameter space shown in Table~\ref{tab:geometry_parameters}.
All geometry variations are introduced through perturbed mesh volume points obtained by warping the baseline reference mesh in accordance with the specified geometry parameters.
Pitch is considered by rotating the entire volume mesh about the y-axis.
Each CFD simulation is run for 1500 solver time steps or until convergence is achieved.
All cases are verified to have a maximum residual tolerance below $1\times10^{-6}$ for all variables. 
For the multifidelity models, panel method and Euler solutions are evaluated at the same geometry and flow condition and mapped to the RANS graph as described in the following sections.

\subsection{Panel and Euler Flow Inputs}
\label{section:methodology:low_fidelity}

We use two low-fidelity flow models to supply the pressure, velocity, and temperature inputs described in Section~\ref{section:methodology:gnn}: a panel method and an Euler CFD model.

\subsubsection{Panel Method}
\label{section:methodology:panel}

We compute panel method solutions using \texttt{VortexAD} \cite{vortexad}, a differentiable solver developed for MDO.
The aircraft surface is discretized into a total of 2,640 triangular and quadrilateral elements, on which surface pressures and velocities are evaluated.
The panel method assumes inviscid, incompressible flow, so it omits the effects of viscosity and compressibility.

To map these predictions to the level-zero graph $G^0$, we first project the baseline graph nodes that lie on the wall boundary onto the baseline panel mesh.
At each projected point, pressure and velocity are interpolated from the contained panel face and its neighboring faces using inverse-distance weighting based on their centroid locations.
We retain these interpolation weights across geometry changes, since the panel mesh and RANS graph have a fixed connectivity.

As it is a surface-based solver, the panel predictions are assigned only at the wall nodes; pressure and velocity at all remaining nodes are assigned freestream values.
Consequently, the additive correction in Eq. \ref{eq:mf} is applied to the freestream state at interior volume nodes, while the panel-method predictions affect the learned correction through message passing from the wall nodes.
In addition, we use the freestream temperature for all graph nodes as the panel method does not predict the temperature field.
Thus, the panel input provides cheap, approximate information about the near-surface flow without requiring calculations in the volume.

\subsubsection{Euler CFD}
\label{section:methodology:euler}

For the Euler-guided model, we solve the steady compressible Euler equations using \texttt{OpenFOAM}.
Simulations use a coarse volume mesh containing 112,176 cells over half of the fluid domain, with the same geometry parameterization and freestream conditions as the RANS cases.
Similar to the RANS model, we accommodate geometry changes by warping the baseline volume mesh while keeping the connectivity fixed.
Unlike the panel method, the Euler model accounts for compressibility, which is important for accurate prediction in the transonic regime.
Although viscous effects are excluded, it provides pressure, velocity, and temperature throughout the volume.
However, this model is more computationally expensive than the panel method, requiring the Euler equations to be solved on a full volume mesh.

We transfer the Euler solution to the RANS graph nodes using a similar inverse distance interpolation approach based on the three closest Euler cell centers for each RANS graph node.
These neighbors are selected using the baseline meshes and retained for all geometries.
For each new geometry, we recompute the interpolation weights from the deformed Euler cell centers and RANS graph node positions, then apply the resulting weights to pressure, velocity, and temperature.

\subsection{Model Implementation and Training}

\begin{table}[t]
\centering
\caption{Multilevel graph hierarchy and message-passing depth during the downsampling and upsampling phases.}
\label{tab:graphs}
\begin{tabular}{ccccccc}
\hline
\textbf{Level} & \textbf{\# nodes} & \textbf{\# edges} & \textbf{avg. edge length} & \textbf{avg. \# cells / cluster} & \textbf{\# down MP layers} & \textbf{\# up MP layers} \\
\hline
$G^0$ & 706k & 2.89M & 3.75 meters & 1.00 & 1 & 1 \\
$G^1$ & 305k & 1.42M & 7.57 meters & 2.31 & 3 & 3 \\
$G^2$ & 119k & 0.54M & 14.81 meters & 5.92 & 5 & --- \\
\hline
\end{tabular}
\end{table}

The GNNs are implemented in JAX\footnote{\url{https://github.com/jax-ml/jax}}, leveraging \texttt{lsdo\_geo} for geometry parameterization, \texttt{VortexAD} for the panel method, \texttt{OpenFOAM} for the Euler solver, and \texttt{IDWarp-JAX}\footnote{\url{https://github.com/LSDOlab/idwarp-jax}} for volume mesh deformation.

For this implementation, we use $L=3$ graph levels and a latent dimension of $d_h=45$.
All learned mappings in the network are implemented as MLPs with two hidden layers and SiLU activation functions.
The encoder, broadcast, and decoder networks use hidden widths of 48, while the message-passing functions $\phi_{\mathrm{msg}}^{(\ell,s)}$ and $\phi_{\mathrm{upd}}^{(\ell,s)}$ use a hidden width of 85.
Details of the resulting graphs are summarized in Table~\ref{tab:graphs}. A single edge in $E^2$ is, on average, four times longer than an edge in $E^0$, allowing messages to propagate farther in the domain per iteration. 

The RANS-GNN, PG-RANS-GNN, and EG-RANS-GNN models are trained separately using the same loss, optimizer, and training schedule.
In Eq.~\ref{eq:loss}, we set $w_{i,k}=4$ for the pressure variable at surface-adjacent volume nodes and $w_{i,k}=1$ otherwise to emphasize surface pressure prediction.
Training uses the AdamW optimizer with a learning rate of $10^{-3}$, lowered to $10^{-4}$ after the validation loss plateaus and then to $10^{-5}$ later in training.
Due to GPU memory limits, the batch size is limited to one geometry per iteration.
Of the 138 converged RANS cases, 110 are used for training, 6 for validation, and 22 for testing with all three models using the same split.
The checkpoint is selected based on a combination of the flow-field loss with reconstructed lift and drag errors on selected validation cases.

\subsection{Differentiable Model Evaluation}
Gradient-based optimization requires derivatives of the objective and constraints with respect to the design variables.
For large-scale aircraft MDO, the aerodynamic quantities involved may be integrated lift, drag, and moment, or surface loads such as those used in structural analysis.
The differentiable GNN model implementations must therefore provide derivatives of the predicted flow field and derived aerodynamic quantities with respect to the model inputs.

As shown in Fig.~\ref{fig:workflow}, geometry changes are first applied to the parameterized aircraft surface using \texttt{lsdo\_geo} and propagated to the surrounding volume graph nodes using \texttt{IDWarp-JAX}.
\texttt{IDWarp-JAX}, a JAX implementation of the inverse-distance-weighted mesh deformation method used in IDWarp \cite{idwarp}, provides derivatives of the volume node coordinates with respect to the surface coordinates and enables efficient GPU-based volume deformations.
Because the mesh connectivity remains fixed, the graph structure and clustering can be reused across geometries, while geometric node and edge features are updated for the deformed geometry.
The flow inputs are then constructed using freestream values for RANS-GNN, mapped panel-method predictions for PG-RANS-GNN, or mapped Euler predictions for EG-RANS-GNN, as described in Section \ref{section:methodology:low_fidelity}.
The single-fidelity GNN predicts the full field directly.
For the multifidelity models, the GNN predicts a correction that is added to the low-fidelity base field.
Pressure forces are computed using the predicted pressures at volume nodes adjacent to the surface, while viscous forces are approximated using Spalding's law of the wall \cite{spalding} along with a small constant additive drag correction term.
The surface forces are then integrated to obtain lift, drag, and pitching moment.
Reconstructing the forces from the reference RANS fields across the training and test datasets yields relative $L_2$ errors in lift and drag coefficients of 0.004\% and 0.084\%, respectively, compared with those computed directly by OpenFOAM.

Derivatives through this workflow are provided using vector--Jacobian products (VJPs), which allow gradient information to be propagated backward through each component of the model.
This allows derivatives of either the predicted flow field or integrated aerodynamic quantities to be connected to other disciplines within the MDO problem.
The geometry parameterization and \texttt{VortexAD} provide derivatives through their \texttt{CSDL}\footnote{\url{https://github.com/LSDOlab/CSDL_alpha}} implementations.
Geometric feature construction, the GNN, and aerodynamic force calculations are differentiated using JAX.
The CSDL and JAX VJPs are chained together through CSDL, allowing derivatives to propagate across the full workflow.
Because the \texttt{OpenFOAM} RANS solutions are used only for training and model analysis, derivatives through the RANS solver are not required.
However, EG-RANS-GNN does require \texttt{OpenFOAM} Euler derivatives during online evaluation, which are unavailable with this current workflow. 
Extending the workflow to support derivatives through the Euler solver is an area for future work.

\section{GNN Performance Analysis}  
\label{section:results}

\subsection{Prediction Accuracy and Computational Cost}

\begin{figure}[htbp]
    \centering
    \includegraphics[width=0.45\textwidth]{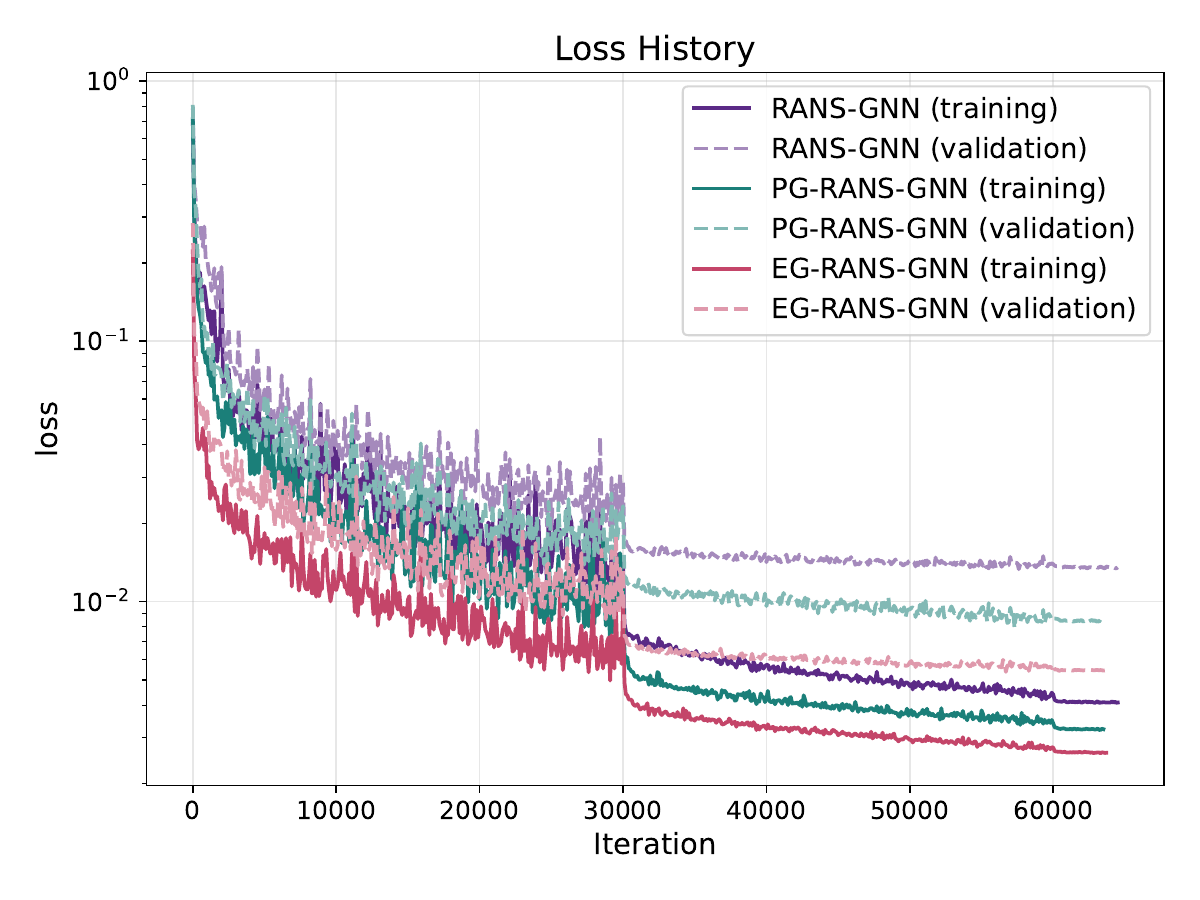}
    \caption{Training and validation loss histories for the RANS-GNN, PG-RANS-GNN, and EG-RANS-GNN models trained on the same dataset. EG-RANS-GNN achieved the lowest validation loss, followed by PG-RANS-GNN and RANS-GNN.}
    \label{fig:loss}
\end{figure}

\begin{table}[htbp]
\centering
\caption{Volume, surface, and aerodynamic coefficient test-set prediction errors. Field quantities report mean relative $L_2$ errors, while aerodynamic coefficients report relative $L_2$ errors across the test set.}
\label{tab:model_accuracy}

\newcolumntype{C}{>{\centering\arraybackslash}p{0.95cm}}
\newcolumntype{D}{>{\centering\arraybackslash}p{1.25cm}}

\begin{tabular}{l
    CCC@{\hspace{8pt}}
    CCC@{\hspace{8pt}}
    DDD}
\toprule
& \multicolumn{3}{c}{Vol. Rel. $L_2$ Error (\%) $\downarrow$}
& \multicolumn{3}{c}{Surf. Rel. $L_2$ Error (\%) $\downarrow$}
& \multicolumn{3}{c}{Coeff. Rel. $L_2$ Error (\%) $\downarrow$} \\
\cmidrule(lr){2-4}
\cmidrule(lr){5-7}
\cmidrule(lr){8-10}

Model
& $p$ & $\mathbf{v}$ & $T$
& $p_s$ & $\mathbf{v}_s$ & $T_s$
& $C_L$ & $C_D$ & $C_{M_{\mathrm{ref}}}$ \\
\midrule

Panel Method
& -- & -- & --
& -- & -- & --
& 22.98 & 72.63 & 23.26 \\

Euler
& -- & -- & --
& -- & -- & --
& 12.40 & 8.34 & 12.49 \\

\midrule

RANS-GNN
& 1.57 & 4.70 & 0.59
& 2.54 & 9.35 & 0.93
& 16.63 & 12.11 & 18.43 \\

PG-RANS-GNN
& 1.17 & 4.02 & 0.49
& 1.96 & 8.15 & 0.82
& 8.21 & 7.07 & 8.58 \\

EG-RANS-GNN
& \textbf{0.82} & \textbf{3.34} & \textbf{0.39}
& \textbf{1.32} & \textbf{6.96} & \textbf{0.67}
& \textbf{5.56} & \textbf{3.02} & \textbf{5.35} \\

\bottomrule
\end{tabular}
\end{table}

We report the predictive accuracy and computational cost of RANS-GNN, PG-RANS-GNN, and EG-RANS-GNN using the same 22 test geometries and taking the RANS solutions for these geometries as the reference fields. 
For comparison, we also show results from the standalone low-fidelity panel method and Euler solver.
Table~\ref{tab:model_accuracy} reports relative $L_2$ errors for pressure, velocity, and temperature in the volume and on the aircraft surface, as well as for the lift, drag, and reference pitching moment coefficients.
We note that the surface-field errors are evaluated at the interior volume nodes adjacent to the aircraft surface, rather than directly on the surface.
The reference pitching moment is computed about a fixed reference point along the aircraft centerline as opposed to the center of gravity and is used here to compare the relative performance of the models.
The corresponding evaluation times and training costs of each model are summarized in Table~\ref{tab:model_computational_cost}.
The training and validation loss histories for all three models are shown in Fig.~\ref{fig:loss}.

Both multifidelity models improve volume-field accuracy over the single-fidelity RANS-GNN for all three predicted quantities.
Adding panel-method-predicted surface fields reduces errors by approximately $14\%$ to $25\%$ compared to RANS-GNN.
EG-RANS-GNN achieves the lowest volume-field errors, reducing them by approximately $29\%$ to $48\%$ compared to RANS-GNN and $17\%$ to $30\%$ compared to PG-RANS-GNN.
This trend is consistent with the additional information provided by the low-fidelity models.
The panel method provides an approximate incompressible and inviscid surface field prediction, while the Euler solver provides a higher-fidelity flow approximation in the volume.

\begin{table}[htbp]
\centering
\caption{Computational performance of the reference solvers and learned models.
CPU timings use 25 MPI ranks. Multifidelity GNN evaluations require both GNN and low-fidelity solver evaluations.
$^{\dagger}$All GNNs require volume graph deformation during end-to-end evaluation, which takes approximately 0.79~s on an RTX~3090 GPU.}
\label{tab:model_computational_cost}

\renewcommand{\arraystretch}{1.12}
\setlength{\tabcolsep}{7.5pt}

\begin{tabular}{@{}lccccc@{}}
\toprule
Model
& Hardware
& Data Gen.
& Training
& End-to-End Eval.
& Speedup vs. RANS \\
\midrule

RANS
& 2$\times$ Xeon Gold 6544Y
& --
& --
& 149.1 s
& 1.00$\times$ \\

Panel Method
& RTX 3090
& --
& --
& \textbf{0.38 s}
& \textbf{392$\times$} \\

Euler
& 2$\times$ Xeon Gold 6544Y
& --
& --
& 15.9 s
& 9.38$\times$ \\

\midrule

RANS-GNN
& RTX 3090
& 5.47 hr
& 24 hr
& 1.10 s$^{\dagger}$
& 136$\times$ \\

PG-RANS-GNN
& RTX 3090
& 5.48 hr
& 24 hr
& 1.48 s$^{\dagger}$
& 101$\times$ \\

\begin{tabular}[c]{@{}l@{}}
EG-RANS-GNN
\end{tabular}
&
\begin{tabular}[c]{@{}c@{}}
2$\times$ Xeon Gold 6544Y \\
RTX 3090
\end{tabular}
& 6.05 hr
& 24 hr
& 16.99 s$^{\dagger}$
& 8.78$\times$ \\

\bottomrule
\end{tabular}
\end{table}

\begin{figure}[htbp]
    \centering
    \includegraphics[width=\textwidth]{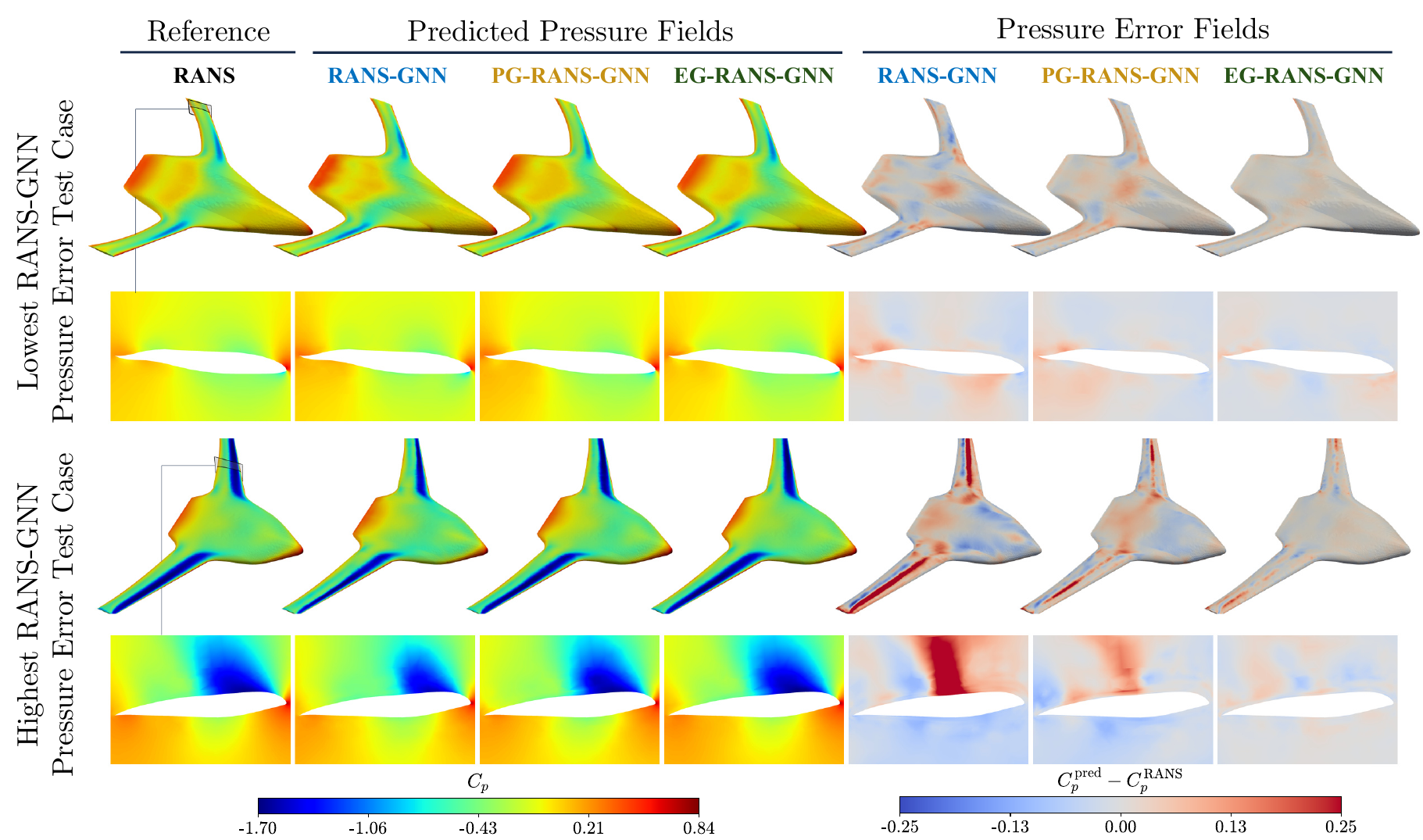}
    \caption{Comparison of surface pressure coefficient predictions and errors for the three GNN models on the test cases with the lowest and highest RANS-GNN pressure prediction errors.
    Both multifidelity models qualitatively reduce pressure prediction errors compared with RANS-GNN, with EG-RANS-GNN providing closer agreement than PG-RANS-GNN.}
    \label{fig:test_fields}
\end{figure}

The same trend is observed for the surface-field predictions, with both multifidelity models improving pressure, velocity, and temperature errors compared to the RANS-GNN.
However, the surface-field errors are higher than the volume-field errors for all three models, potentially due to stronger flow gradients near the surface.
Figure~\ref{fig:test_fields} illustrates differences in predicted pressures using the two test geometries with the smallest and largest RANS-GNN volume pressure prediction errors.
All three models capture the general trends of the RANS solution, although the error plots show that the single-fidelity GNN has larger errors in locations with stronger pressure gradients, such as near the upper surface.
The PG-RANS-GNN model generally lowers these errors, while the EG-RANS-GNN model provides the closest agreement for the two examples shown.

\begin{figure}[htbp]
    \centering
    \includegraphics[width=\textwidth]
    {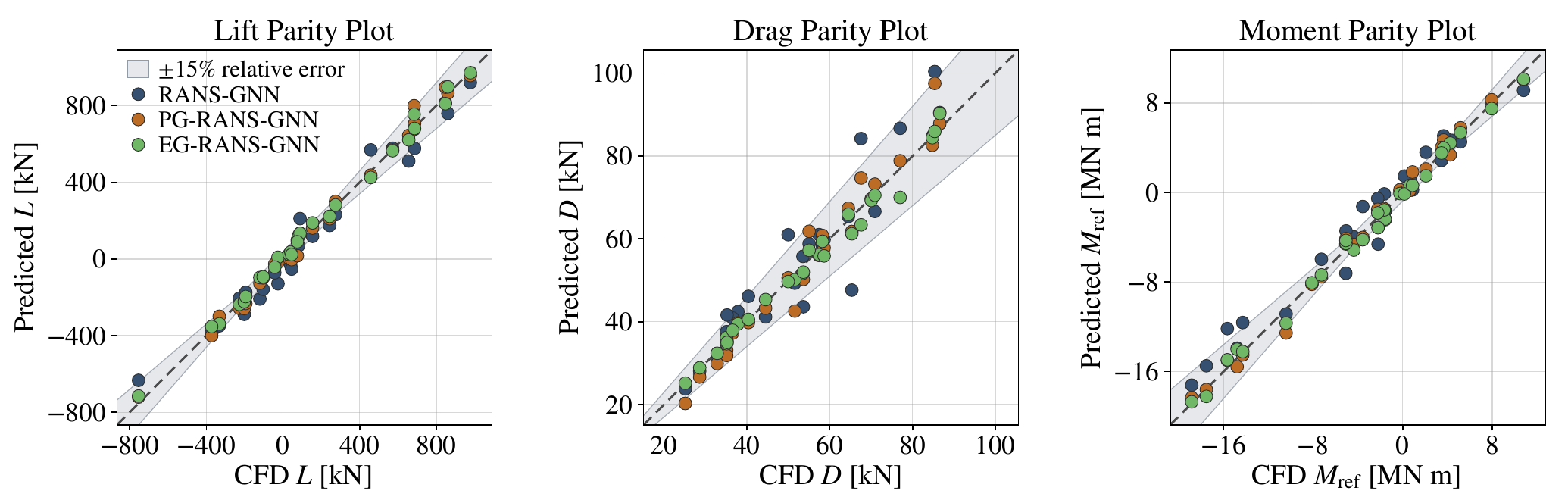}
    \caption{Comparison of predicted lift, drag, and reference pitching moment against reference RANS results for the three GNN models on the test dataset. Both multifidelity GNNs show closer agreement with RANS CFD than RANS-GNN, with EG-RANS-GNN achieving the lowest prediction errors for all three.}
    \label{fig:model_parity}
\end{figure}

We also compute lift, drag, and pitching moment about a reference point from the predicted surface fields to assess the quality of the aerodynamic quantities in addition to the field predictions.
These integrated quantities are not explicitly included in the training loss and they are computed from the learned fields.
Relative to RANS-GNN, PG-RANS-GNN reduces the relative $L_2$ errors in lift, drag, and pitching moment coefficients by approximately 51\%, 42\%, and 53\%, respectively.
EG-RANS-GNN achieves even larger reductions of approximately 67\%, 75\%, and 71\% for the same quantities.
Both multifidelity GNNs outperform their corresponding standalone low-fidelity solvers on all three reported coefficient errors.
The parity plots in Fig.~\ref{fig:model_parity} show individual errors for all test geometries, where it can be seen that the multifidelity models generally provide closer agreement with RANS CFD across a range of drag, lift, and moment values. 
The GNN models especially improve drag prediction compared to the panel method, which neglects viscous and compressibility effects.

\begin{figure}[htbp]
    \centering
    \includegraphics[width=0.55\textwidth]
    {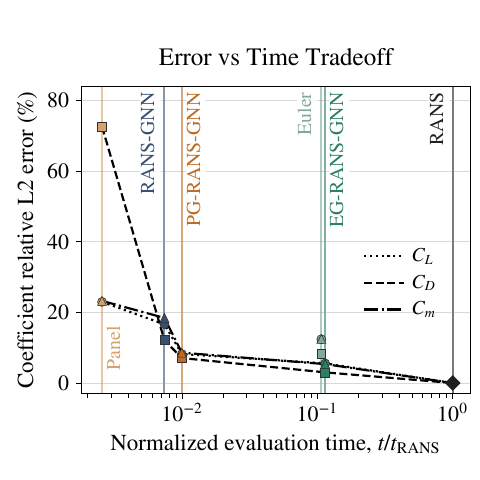}
    \caption{Tradeoff between aerodynamic coefficient relative $L_2$ error and evaluation time normalized by the RANS evaluation time, with lines indicating the computed Pareto fronts. Increasing the fidelity of the GNN inputs generally improves prediction accuracy at the expense of computational cost.}
    \label{fig:pareto_coefficients}
\end{figure}

Table~\ref{tab:model_computational_cost} compares the computational costs of the reference solvers and the three GNN models, including the necessary mesh deformation and low-fidelity evaluations.
For all three GNN models, deformation of the volume graph using \texttt{IDWarp-JAX} takes approximately $0.79\,\mathrm{s}$, while feature construction and GNN inference require only $0.31\,\mathrm{s}$ both on an RTX 3090 GPU using single precision.
Thus, for PG-RANS-GNN and RANS-GNN, the deformation of the graph accounts for most of the evaluation cost, while the Euler solve dominates the evaluation cost of EG-RANS-GNN.
Despite these extra steps, EG-RANS-GNN is around nine times faster than the RANS CFD, while RANS-GNN and PG-RANS-GNN achieve speedups of roughly two orders of magnitude.
The GPU-accelerated panel method remains the fastest approach, primarily because it does not require the expensive mesh deformation.
Unlike the physics-based solvers, the GNN models also have additional offline costs for data generation and training, which must be taken into account.
For training, all GNN preprocessing steps, including the low-fidelity solves, are performed in advance, meaning each iteration requires only the GNN backward pass, which takes approximately 0.99 s.
We also note that the reported data generation times do not include failed CFD runs and may therefore underestimate the total dataset generation cost.
Finally, these timings represent wall clock timing comparisons between the models and can vary with hardware and implementation.

Figure~\ref{fig:pareto_coefficients} shows the tradeoff between accuracy and evaluation time across the different models.
As expected, increasing GNN input fidelity generally improves prediction accuracy at the expense of additional computational cost.
The choice of model for optimization therefore depends on the accuracy requirements of the problem, with the various approaches offering advantages for different applications.

Overall, the results show that providing the neural networks with physics-based low-fidelity information can improve both the field and coefficient predictions compared to RANS-GNN.
The level of improvement depends on the fidelity of the inputs, although it comes at the cost of additional evaluation time.

\subsection{Training-Data Efficiency} \label{section:results:ablation}
\begin{figure}[htbp]
    \centering
    \includegraphics[width=\textwidth]{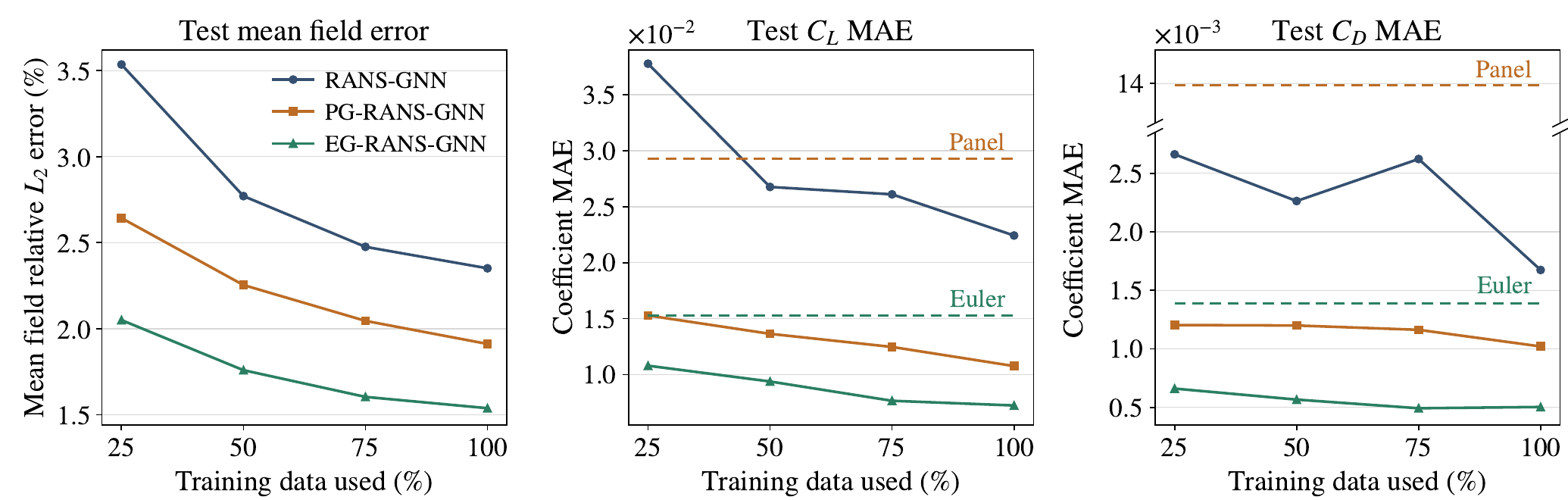}
    \caption{Effect of high-fidelity training dataset size on mean flow-field prediction error and lift and drag coefficient mean absolute errors for the three GNN models. Both multifidelity models outperform RANS-GNN across all training dataset sizes, with PG-RANS-GNN and EG-RANS-GNN requiring only 50\% and 25\% of the full training dataset (110 RANS cases), respectively, to achieve lower mean flow-field errors than RANS-GNN trained on the full dataset.}
    \label{fig:three_panel_ablation_figure_hardcoded}
\end{figure}

To evaluate the effect of training data availability, we train each model using 25\%, 50\%, 75\%, and 100\% of the full original training dataset. 
The smaller datasets are built as subsets of the larger ones to ensure new data is not introduced when decreasing the size. 
As expected, Fig.~\ref{fig:three_panel_ablation_figure_hardcoded} shows that the relative $L_2$ error averaged across the five output states decreases consistently with increasing training data for all three models.
Both multifidelity models outperform RANS-GNN across all training set sizes, with the largest improvements occurring when only 25\% of the available training data is used.
Notably, EG-RANS-GNN trained with only 25\% of the data achieves a lower mean field error than RANS-GNN trained on the full dataset.
Further, PG-RANS-GNN trained with 25\% of the data achieves a lower mean field error than RANS-GNN trained on 50\% of the dataset.

The aerodynamic coefficient errors show similar overall trends, although the improvements with increasing training data are less consistent than for the field predictions.
Even with only 25\% of the training data, PG-RANS-GNN achieves lower lift and drag coefficient errors than RANS-GNN trained on the full dataset, while EG-RANS-GNN trained on 25\% of the data achieves comparable or lower errors than PG-RANS-GNN trained on the full dataset.
Although not shown, the pitching moment coefficient errors follow trends similar to those observed for lift.
Increasing the training set size also generally leads to lower improvements in drag error compared to lift and field errors for all the networks.
Overall, these results suggest that using the multifidelity GNN method can reduce the amount of high-fidelity training data needed to achieve comparable prediction accuracy to the single-fidelity GNN, which is useful when training data is limited or expensive to generate.

\subsection{Aerodynamic Trends and Sensitivities} \label{section:results:trends}

\begin{figure}[htbp]
    \centering
    \includegraphics[width=\textwidth]
    {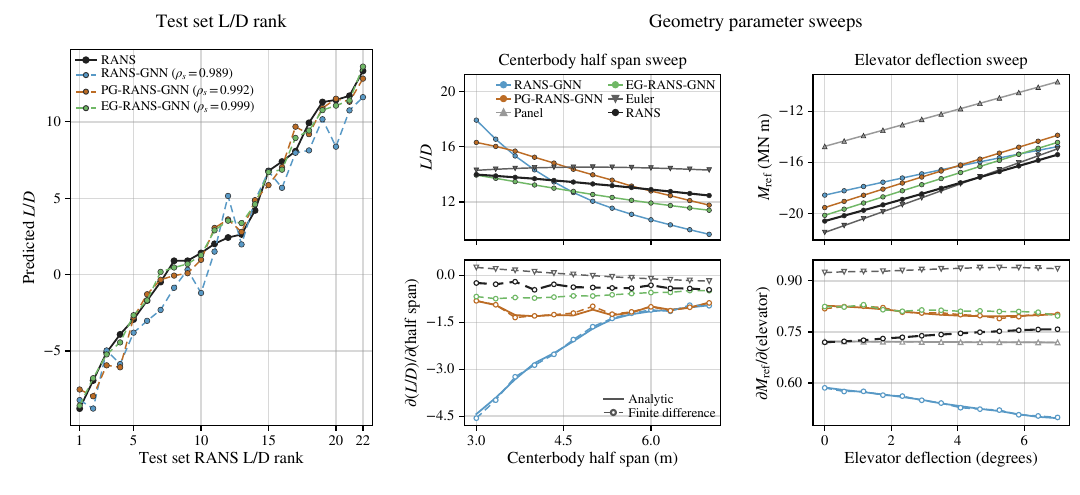}
    \caption{Test-set lift-to-drag ratio ranking and aerodynamic
    response to centerbody half-span and elevator-deflection
    variations, including analytical and finite-difference
    sensitivity comparisons where available; the Euler and
    RANS CFD reference sensitivities are computed using
    finite differences only.}
    \label{fig:ranking_derivative_sweeps}
\end{figure}

In addition to forward prediction accuracy, we also investigate how well the models capture trends and sensitivities, both of which are important for gradient-based optimization.
Figure~\ref{fig:ranking_derivative_sweeps} shows that all three GNNs mostly preserve the $L/D$ ranking of the 22 test geometries relative to the RANS reference, with the multifidelity models showing closer agreement than RANS-GNN.
We also show two representative parameter sweeps about the baseline geometry which is not included in the training set: $L/D$ versus centerbody half-span and pitching moment versus elevator deflection.
All models capture the general trends compared to the reference, although differences can be seen in their predicted sensitivities.
In both examples, PG-RANS-GNN appears to provide derivatives that qualitatively agree more closely with the RANS finite-difference results compared to those of RANS-GNN.
The panel-method and Euler sensitivities are also smoother compared to the GNN derivatives, which may be due to the fact that single precision is used for the neural network calculations.
The analytical derivatives are also generally consistent with the finite-difference approximations of the same model, showing that the derivatives are implemented correctly.
Finally, we note again that the RANS and Euler solvers have only the finite difference approximations shown here, due to the unavailability of derivatives through the CFD solver.

\section{Demonstration of GNN-Based Aircraft Design Optimization} \label{section:optimization}

\subsection{Camber and Twist Optimization}\label{sec:camber_opt}

\begin{figure}[htbp]
    \centering
    \includegraphics[width=\textwidth]{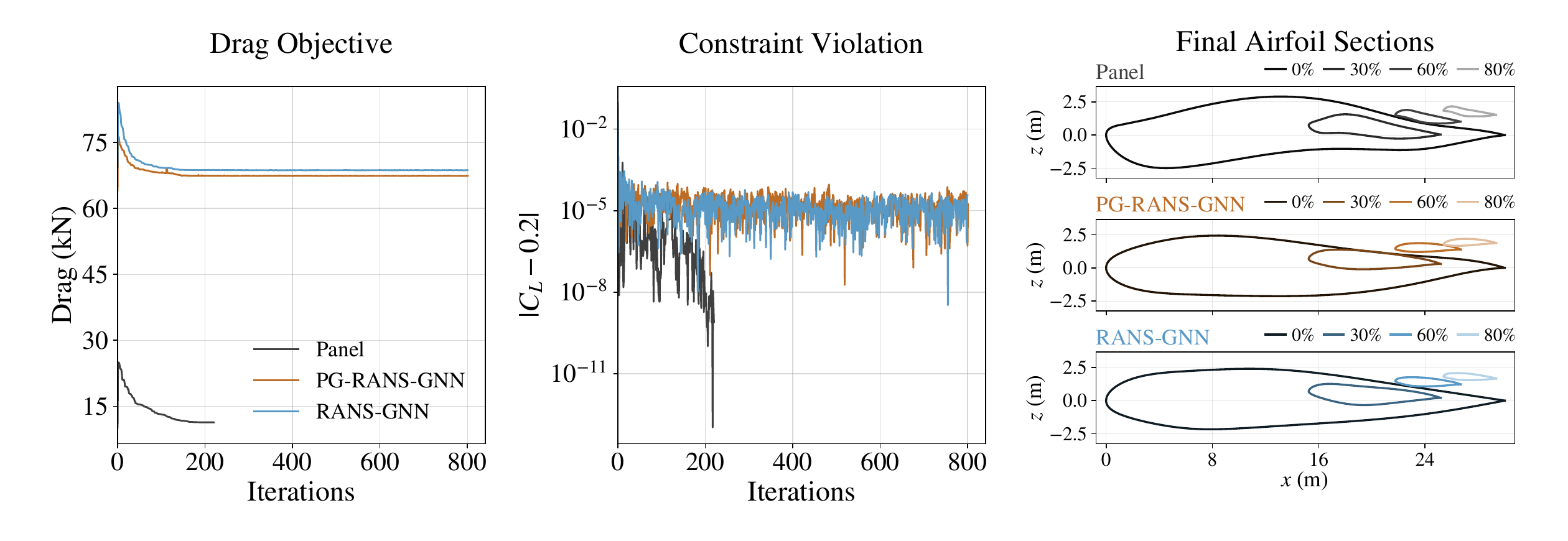}
    \caption{Predicted drag objective, lift-constraint residual, and final airfoil sections at four spanwise stations for each model. All three models reduce their predicted drag while approaching the target lift coefficient.}
    \label{fig:camber_optimization_history}
\end{figure}

To demonstrate the use of the surrogate models in gradient-based optimization, we first consider an aerodynamic shape optimization problem.
This problem minimizes drag at cruise subject to a fixed lift coefficient of 0.2, which the baseline geometry does not satisfy.
The problem contains 52 design variables, consisting of 48 B-spline surface control points parameterizing airfoil camber and 4 twist variables along the wing, all drawn from the geometry parameters listed in Table \ref{tab:geometry_parameters}.
The remaining geometry parameters, including the planform variables, are fixed at the baseline values.
We solve the problem using RANS-GNN, PG-RANS-GNN, and the panel method to compare their behavior during optimization.
The optimization problem is solved using the interior-point algorithm implemented in \texttt{modOpt} \cite{modopt}.

Figure~\ref{fig:camber_optimization_history} shows the objective and constraint history as well as the final optimized airfoil sections for the three models.
All models reduce predicted drag and satisfy the predicted lift constraint, demonstrating their applicability in gradient-based optimization.
However, we note that the objective and constraint values reported by each model are approximations.

\begin{table}[htbp]
\centering
\small
\caption{Comparison of predicted aerodynamic coefficients with
RANS re-evaluations for the three optimized geometries.
Relative errors are computed with respect to the corresponding
RANS values.}
\label{tab:camber_opt_accuracy}

\setlength{\tabcolsep}{5pt}
\renewcommand{\arraystretch}{1.15}

\begin{tabular}{l
    ccc@{\hspace{18pt}}
    ccc@{\hspace{18pt}}
    ccc}
\toprule
Model
& $C_L$ & $C_L^{\mathrm{RANS}}$ & Err. (\%)
& $C_D$ & $C_D^{\mathrm{RANS}}$ & Err. (\%)
& $C_{M_{\mathrm{ref}}}$ & $C_{M_{\mathrm{ref}}}^{\mathrm{RANS}}$ & Err. (\%) \\
\midrule

Panel Method (Optimized)
& 0.2001 & 0.1508 & 32.69
& 0.0019 & 0.0402 & 95.27
& -0.2306 & -0.1648 & 39.93 \\

RANS-GNN (Optimized)
& 0.2005 & 0.1818 & 10.29
& 0.0110 & 0.0132 & 16.67
& -0.2446 & -0.2301 & 6.30 \\

PG-RANS-GNN (Optimized)
& 0.2001 & 0.1935 & \textbf{3.41}
& 0.0108 & 0.0126 & \textbf{14.29}
& -0.2553 & -0.2519 & \textbf{1.35} \\

\bottomrule
\end{tabular}
\end{table}

\begin{figure}[htbp]
    \centering
    \includegraphics[width=\textwidth]{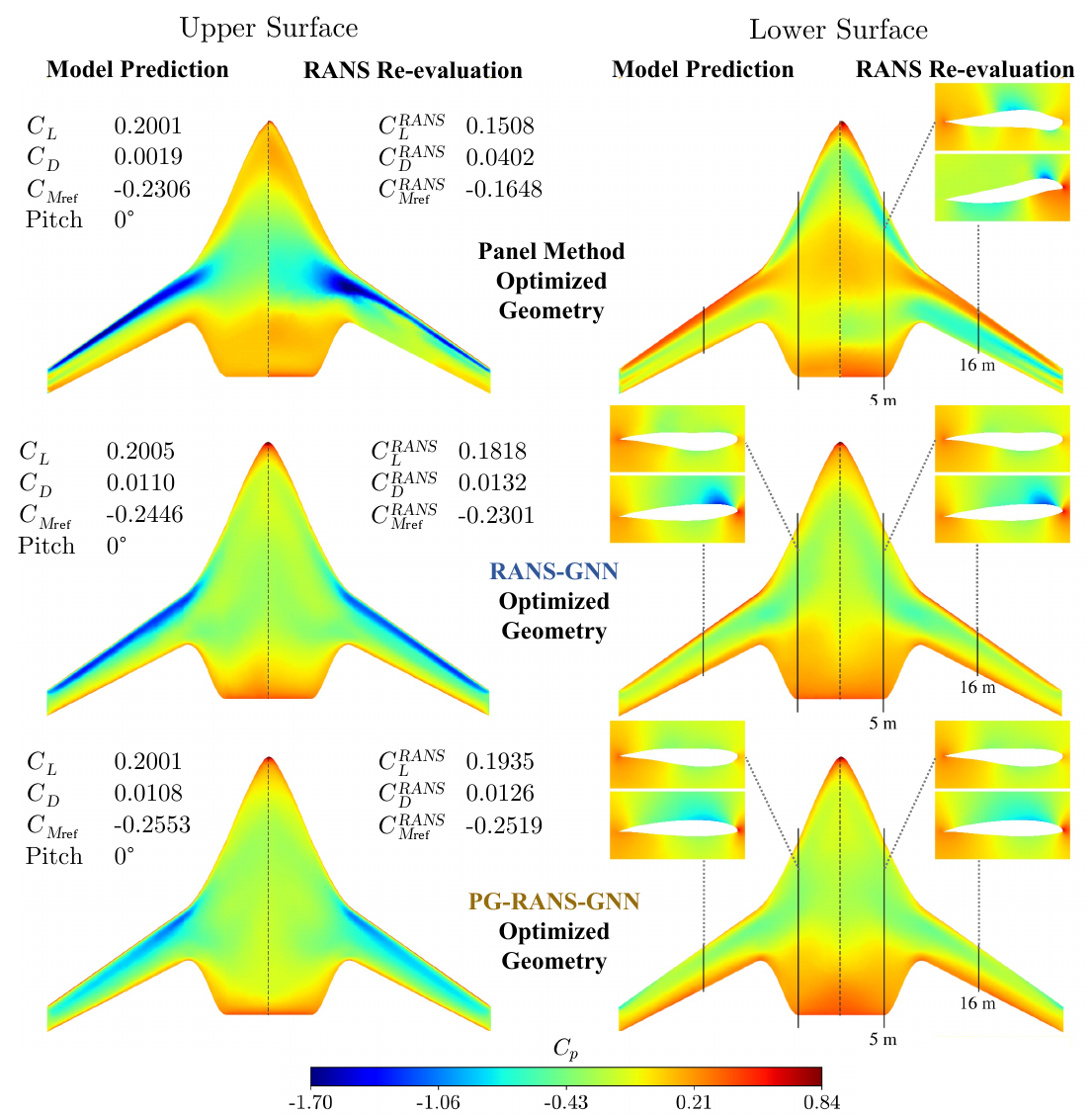}
    \caption{Comparison of predicted pressure coefficient distributions and RANS CFD re-evaluations for the three geometries obtained from the camber optimization problem.}
    \label{fig:final_camber_opt}
\end{figure}

The optimized airfoil sections in Fig.~\ref{fig:camber_optimization_history} show that the two GNN-based optimizations produce relatively similar geometries, while the panel method leads to more aggressive changes in camber.
Figure~\ref{fig:final_camber_opt} compares the predicted pressure distributions for the optimized geometries with RANS CFD re-evaluations.
Both GNN models capture the general pressure distribution trends for both the surface and the cross sections.
In contrast, the panel-method-optimized geometry shows larger discrepancies in the pressure distribution, particularly near the upper and lower surfaces of the wing.
These discrepancies may reflect viscous and compressibility effects not captured by the panel method.

We evaluate the final optimized geometries using the RANS CFD solver.
Table~\ref{tab:camber_opt_accuracy} compares the predicted aerodynamic coefficients with the RANS re-evaluations of each optimized geometry.
PG-RANS-GNN achieves the lowest relative errors for lift, drag, and pitching moment, showing better agreement with RANS than either RANS-GNN or the standalone panel method.
In contrast, the panel-method optimization leads to the largest errors, particularly for drag, while achieving the lowest predicted objective during optimization.
Overall, these results demonstrate that both GNN-based models can be used for gradient-based aerodynamic optimization, while incorporating low-fidelity information improves the accuracy of the aerodynamic predictions at the optimized design.

\subsection{Multidisciplinary Optimization for Fuel-Burn Minimization}\label{sec:large_scale_mdo}

\begin{figure}[htbp]
    \centering
    \includegraphics[width=\textwidth]{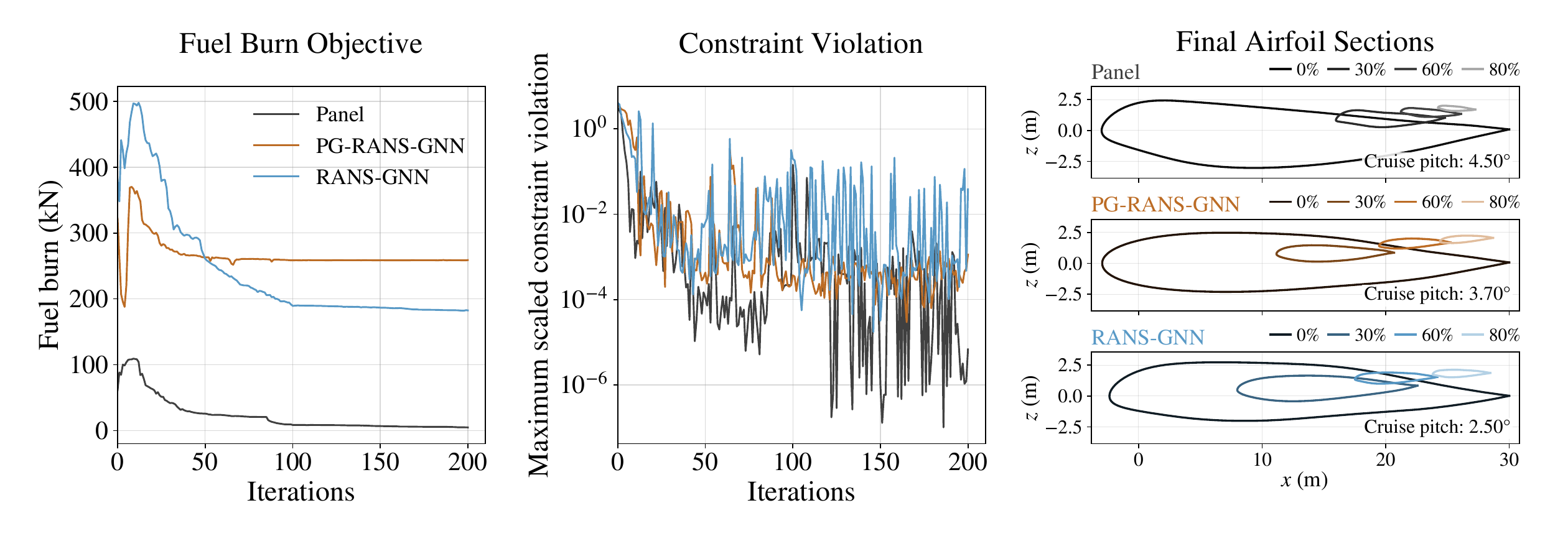}
    \caption{Predicted fuel-burn objective, maximum predicted constraint violation, and final airfoil sections at four spanwise stations for each model. All three models reduce their predicted objective while reducing predicted constraint violation.}
    \label{fig:mdo_optimization_history}
\end{figure}

\begin{table}[htbp]
\centering
\small
\caption{Variables and constraints in the MDO problem with a fuel burn objective.}
\label{tab:mdo_problem}
\setlength{\tabcolsep}{6pt}
\begin{tabularx}{\textwidth}{@{}Xr@{\hspace{2em}}Xr@{}}
\toprule
Name & Qty & Name & Qty \\
\midrule
\multicolumn{4}{@{}l}{\textbf{Design variables}} \\
Planform and airfoil geometry\textsuperscript{$\dagger$} & 129
  & Cruise pitch & 1 \\
Structural-sizing pitch & 1
  & Cruise elevator deflection & 1 \\
Beam top thicknesses & 10
  & Beam web thicknesses & 10 \\
\textbf{Total design variables} & \textbf{152} & & \\
\midrule
\multicolumn{4}{@{}l}{\textbf{Constraints}} \\
Transition leading-edge monotonicity & 4
  & Cruise force balance & 1 \\
Structural-sizing force balance & 1
  & Cruise moment trim & 1 \\
Maximum beam stress & 1
  & Fuselage volume & 1 \\
Fuel storage volume & 1
  & Takeoff and landing lengths & 2 \\
Static margin & 1
  & & \\
\textbf{Total constraints} & \textbf{13} & & \\
\bottomrule
\end{tabularx}

\vspace{3pt}
\begin{minipage}{\textwidth}
\footnotesize
\textsuperscript{$\dagger$}The 129 geometry variables use the parameterization and sampling ranges in Table~\ref{tab:geometry_parameters},
except for the airfoil thickness and camber coefficients, whose optimization bounds are slightly tighter. Pitch and elevator
deflection for the cruise condition are listed separately here.
\end{minipage}
\end{table}

Next, we solve a full large-scale multidisciplinary design optimization (MDO) problem.
We adapt the problem defined in Scotzniovsky et al. \cite{scotzniovsky2026large}, which aims to minimize fuel burn at cruise while taking into account trim, structural sizing, stability, and geometry constraints.
Unlike the previous camber optimization, this problem allows for changes in the aircraft planform, pitch, and elevator deflection, in addition to the airfoil shapes.
Further, the problem relies on multiple surrogate model evaluations instead of one to compute the stability constraints.
Finally, we note that the same interior-point optimizer as in Section \ref{sec:camber_opt} is used.

The optimization problem consists of 152 design variables and 13 scalar constraints, as summarized in Table~\ref{tab:mdo_problem}, and includes a nominal cruise condition, a structural-sizing condition, and a stability condition evaluated by perturbing the cruise pitch.
The design variables include 130 aircraft geometry parameters as tabulated in Table~\ref{tab:geometry_parameters}, cruise pitch, a separate structural-sizing pitch variable, along with 20 structural thickness parameters, each bounded between $1$ and $100~\mathrm{mm}$.
At the cruise condition, the aerodynamic model provides lift, drag, and pitching moment predictions, which are then used to compute the fuel burn objective and enforce lift and moment trim constraints.
Fuel burn is estimated using the Breguet range equation based on the predicted lift-to-drag ratio, aircraft weight, cruise range, and thrust-specific fuel consumption.
For the structural sizing condition, surface forces are computed using the panel method at a +2.5g maneuver condition and transferred to a beam model to estimate structural weight and stresses.
The maximum beam stress is constrained to $324~\mathrm{MPa}$ after applying a 1.5 safety factor.
We note that as the structural sizing condition is not at cruise, the panel method is used to compute structural forces for all three optimization variants.
Longitudinal stability is computed by perturbing the cruise pitch and evaluating the model at the perturbed condition to enforce a minimum static margin constraint of 2\%.
Takeoff and landing distances are estimated from the aircraft weight, wing area, and engine thrust with each constrained to be below \(10{,}000\,\mathrm{ft}\).
The fuselage volume computed from the geometry must be at least \(1{,}183.6\,\mathrm{m}^3\).
The usable fuel storage, which is estimated as half the transition section volume, must accommodate the required cruise fuel volume plus a 20\% reserve.
Finally, we add a constraint to enforce monotonic sweep at the leading edge of the transition section.

Figure~\ref{fig:mdo_optimization_history} shows the fuel-burn objective, maximum scaled constraint violation, and optimized airfoil sections for the three optimizations.
All three models reduce their predicted fuel-burn objectives from the baseline, although the optimization histories and final objective values differ significantly between them.
In particular, the panel method predicts a much lower fuel burn than either GNN model due to significant drag underprediction, while the RANS-GNN and PG-RANS-GNN models predict closer final fuel-burn values.
We can also observe that all three models predict feasible designs with only minor constraint violations, although the RANS-GNN model displays larger constraint violation fluctuations in the optimization history compared to the other two models.
We also note that the final objective and constraint values do not necessarily reflect the ground truth values due to model prediction errors.

\begin{figure}[htbp]
    \centering
    \includegraphics[width=\textwidth]{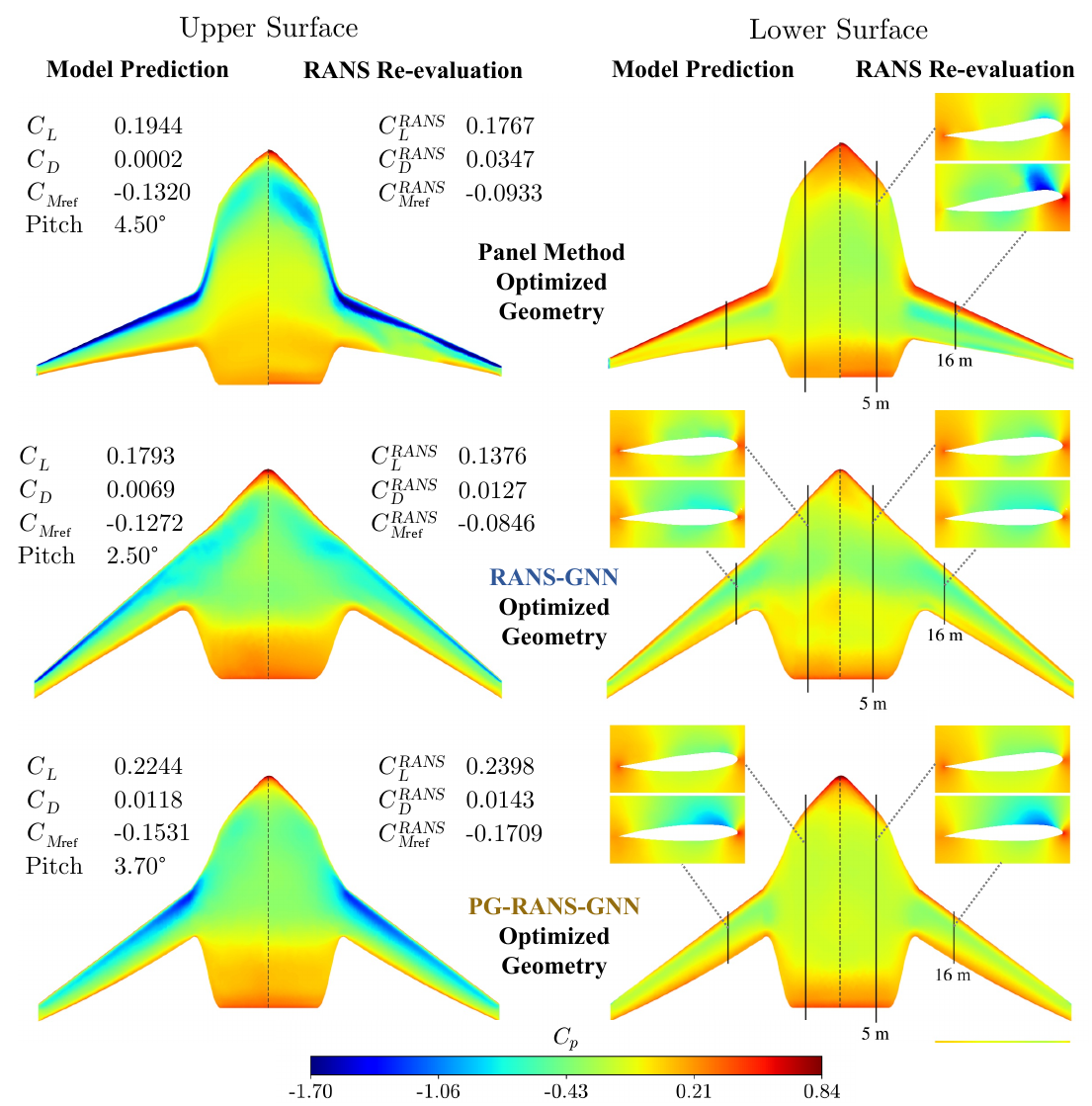}
    \caption{Comparison of predicted pressure coefficient distributions and RANS CFD re-evaluations for the three geometries obtained from the full MDO problem.}
    \label{fig:final_mdo_opt}
\end{figure}

Figure~\ref{fig:final_mdo_opt} compares the optimized geometries and their predicted pressure distributions along with RANS CFD re-evaluated on the final designs.
The three optimizations produce significantly different planform designs and airfoil shapes.
The panel method optimization leads to the design with the highest cruise pitch angle, the most aft wing, and the highest sweep angle. 
In contrast, the RANS-GNN design has the opposite characteristics, with the lowest cruise pitch angle, most forward wing, and lowest sweep angle.
Finally, the PG-RANS-GNN design lies between the other two in terms of pitch and wing position.
The cruise centerbody elevator deflection follows the opposite trend to the pitch angle to maintain moment trim.
The final airfoil shapes have similar trends to the camber-only optimization, with the panel method design having aggressive negative camber.
All three designs increased their overall planform areas to $725~\mathrm{m^2}$, $901~\mathrm{m^2}$, and $766~\mathrm{m^2}$ for the panel method, RANS-GNN, and PG-RANS-GNN, respectively, compared to the baseline area of $529~\mathrm{m^2}$.
This is consistent with a $10\%$ increase in the centerline chord for all models along with an increase of $25.7\%$, $38.9\%$ and $26.4\%$ for total tip-to-tip span for the panel method, RANS-GNN, and PG-RANS-GNN, respectively.

Both GNN models capture the general RANS pressure distribution trends across the upper and lower surfaces, although there are visible differences in the low pressure region on the upper surface near the leading edge for the RANS-GNN case.
PG-RANS-GNN shows better qualitative agreement with the RANS surface pressure distributions, although some differences remain in the volumetric pressure fields above the upper surface.
In comparison, the panel method shows larger discrepancies, particularly on the middle section of the wing upper surface.
This is expected as the panel method does not capture viscous and compressibility effects.
Overall, the pressure comparisons show that the PG-RANS-GNN provides the closest qualitative agreement with RANS CFD at its optimized geometry.

Although the predicted fields capture general trends, differences remain in the aerodynamic coefficients, as summarized in Table~\ref{tab:mdo_opt_accuracy}.
PG-RANS-GNN achieves the lowest relative errors across all three variables, showing better agreement with RANS CFD than the single-fidelity RANS-GNN.
Interestingly, the panel method predicts lift and reference pitching moment more accurately than RANS-GNN, despite its much larger drag error.
Although the previous camber-only optimization and the analysis on the test dataset showed RANS-GNN outperforming the panel method, the large-scale MDO results suggest that the optimizer can exploit inaccuracies in the surrogate model, resulting in a potentially less accurate prediction at the optimized design.
In this case, the RANS-GNN overpredicts lift and underpredicts drag, compared to the RANS re-evaluation.
Overall, this comparison suggests that incorporating low-fidelity physics into the GNN can improve the reliability of its aerodynamic predictions at the optimized geometry, even within a large-scale MDO problem.

\begin{table}[htbp]
\centering
\small
\caption{Comparison of predicted aerodynamic coefficients with
RANS re-evaluations for the three MDO-optimized geometries.
Relative errors are computed with respect to the corresponding
RANS values.}
\label{tab:mdo_opt_accuracy}

\setlength{\tabcolsep}{5pt}
\renewcommand{\arraystretch}{1.15}

\begin{tabular}{l
    ccc@{\hspace{10pt}}
    ccc@{\hspace{10pt}}
    ccc}
\toprule
Model
& $C_L$ & $C_L^{\mathrm{RANS}}$ & Err. (\%)
& $C_D$ & $C_D^{\mathrm{RANS}}$ & Err. (\%)
& $C_{M_{\mathrm{ref}}}$ & $C_{M_{\mathrm{ref}}}^{\mathrm{RANS}}$ & Err. (\%) \\
\midrule

Panel Method (Optimized)
& 0.1944 & 0.1767 & 10.02
& 0.0002 & 0.0347 & 99.42
& -0.1320 & -0.0933 & 41.48 \\

RANS-GNN (Optimized)
& 0.1793 & 0.1376 & 30.31
& 0.0069 & 0.0127 & 45.67
& -0.1272 & -0.0846 & 50.35 \\

PG-RANS-GNN (Optimized)
& 0.2244 & 0.2398 & \textbf{6.42}
& 0.0118 & 0.0143 & \textbf{17.48}
& -0.1531 & -0.1709 & \textbf{10.42} \\

\bottomrule
\end{tabular}
\end{table}

\section{Conclusion} \label{section:significance}
In this paper, we presented a multifidelity graph neural network (GNN) method for 3D aerodynamic flow prediction and applied it to RANS analysis around a blended-wing-body aircraft concept in a workflow suitable for gradient-based optimization.
The method uses a low-fidelity solver to provide an approximate flow field which serves as both the GNN input and a baseline prediction that the GNN corrects towards the high-fidelity RANS flow field.

The results for the panel-method-guided and Euler-CFD-guided multifidelity models show that incorporating low-fidelity aerodynamic information improves both flow-field and aerodynamic-coefficient predictions compared with the single-fidelity GNN.
Further, increasing the fidelity of the base model improves the multifidelity GNN's prediction accuracy at the expense of additional computational cost.
However, the panel-guided and Euler-guided GNNs achieved end-to-end speedups of approximately 100 times and 9 times over RANS CFD, respectively.
The panel-guided and Euler-guided GNNs, trained on only 50\% and 25\% of the RANS training data, respectively, achieved lower mean flow-field errors than the single-fidelity GNN trained on the full dataset.
Finally, in the optimization demonstrations, the panel-guided GNN model provided more accurate aerodynamic predictions at its optimized geometries than the single-fidelity GNN and standalone panel method solver, based on RANS re-evaluations.

Several important directions remain for future work.
First, enabling derivatives through the RANS and Euler CFD solvers would allow additional optimization comparisons and enable end-to-end derivatives for the Euler-guided multifidelity GNN.
A more extensive ablation study is also needed to assess the contributions of the low-fidelity inputs, graph features, and network architecture, along with a more systematic investigation of model hyperparameters.
Further, evaluating the method on larger, well-validated datasets would strengthen confidence in our model for realistic optimization problems.
Expanding the training dataset to include additional flight conditions would enable use of the method beyond the one cruise condition considered here.
Finally, using the multifidelity GNNs to accelerate coupled optimization problems such as aerostructural cases is an important direction for future work.

\section*{Acknowledgments}
The authors would like to acknowledge the Transdisciplinary Design Methods and Transdisciplinary Design Exploration branches of the Air Warfare Directorate, Air Force Research Laboratory for funding and supporting this effort through the Collaborative Center for Design and Research of Interdisciplinary Systems program.

The authors would also like to thank Edward Lowell for his assistance with CFD, as well as Jason Kao for his regular feedback.

The authors used OpenAI’s ChatGPT and Codex to improve manuscript language and readability, as well as to assist with parts of the code implementation.
The authors take full responsibility for the content, methodology, and results presented in this work.

Distribution A: Approved for public release; distribution is unlimited. AFRL-2026-2025.

\bibliography{sample}

\end{document}